\documentclass[letterpaper]{article} 
\usepackage[submission]{aaai24}  
\usepackage{times}  
\usepackage{helvet}  
\usepackage{courier}  
\usepackage[hyphens]{url}  
\usepackage{graphicx} 
\usepackage{float}
\usepackage{multirow}
\usepackage{multicol}
\usepackage{caption}
\usepackage{tcolorbox}
\usepackage{amsmath}
\usepackage{amsfonts} 
\usepackage{listings}
\usepackage{xcolor}
\usepackage{multicol}
\usepackage{soul}

\makeatletter
\NewDocumentCommand{\sotwo}{O{gray}O{black}+m}
    {%
        \begingroup
        \setulcolor{#1}%
        \setul{-.7ex}{2.2pt}%
        \def\SOUL@uleverysyllable{%
            \rlap{%
                \color{#2}\the\SOUL@syllable
                \SOUL@setkern\SOUL@charkern}%
            \SOUL@ulunderline{%
                \phantom{\the\SOUL@syllable}}%
        }%
        \ul{#3}%
        \endgroup
    }
\makeatother

\definecolor{lightestgray}{rgb}{0.95, 0.95, 0.95}

\lstdefinestyle{basicStyle1}{
  language=Java,
  basicstyle=\ttfamily\small,
  numbers=none,
  breaklines=true,
  captionpos=b,
  frame=single,
  backgroundcolor=\color{lightestgray}, 
  rulecolor=\color{black}
}

\usepackage{natbib}  
\usepackage{caption} 
\usepackage{algorithm}
\usepackage{algorithmic}
\usepackage{booktabs}
\usepackage{newfloat}
\usepackage{listings}
\DeclareCaptionStyle{ruled}{labelfont=normalfont,labelsep=colon,strut=off} 
\floatstyle{ruled}
\newfloat{listing}{tb}{lst}{}
\floatname{listing}{Listing}
\usepackage{xcolor}

\title{Can LLMs Identify Gaps and Misconceptions in Students' Code Explanations?}
\author{
    Priti Oli,
    Rabin Banjade,
    Andrew M. Olney,
    Vasile Rus
}
\affiliations{
     \{poli,rbnjade1,aolney,vrus\}@memphis.edu \\
    \textsuperscript{\rm }University of Memphis\\

       3720 Alumni Ave, Memphis, TN 38152\\
}

\begin{document}

\maketitle

\begin{abstract}

This paper investigates various approaches using Large Language Models (LLMs) to identify gaps and misconceptions in students' self-explanations of specific instructional material, in our case explanations of code examples. This research is a part of our larger effort to automate the assessment of students' freely generated responses, focusing specifically on their self-explanations of code examples during activities related to code comprehension. In this work, we experiment with zero-shot prompting, Supervised Fine-Tuning (SFT), and preference alignment of LLMs to identify gaps in students' self-explanation. With simple prompting, GPT-4 consistently outperformed LLaMA3 and Mistral in identifying gaps and misconceptions, as confirmed by human evaluations. Additionally, our results suggest that fine-tuned large language models are more effective at identifying gaps in students' explanations compared to zero-shot and few-shot prompting techniques. Furthermore, our findings show that the preference optimization approach using Odds Ratio Preference Optimization (ORPO) outperforms SFT in identifying gaps and misconceptions in students' code explanations. 


\end{abstract}

\section{Introduction}

Large Language Models (LLMs), such as ChatGPT, have captured the attention of many researchers due to their remarkable ability to generate responses to user prompts and their competitive performance on varied tasks such as question answering, summarization, and semantic similarity~\cite{zhao2023survey}. In the context of education, LLMs offer opportunities to customize learning experiences and adapt instructional material according to the unique needs of each learner~\cite{giannakos2024promise}. Such AI-powered adaptive learning systems have been explored to analyze student data~\cite{oli2023automated}, identify learning gaps~\cite{banjade2024identifying}, and to tailor instructional material resulting in enhanced engagement and improved learning outcomes~\cite{oli2023improving}. While LLMs provide immediate and round-the-clock support, it is crucial that they consistently exhibit fairness, accuracy, and reliability~\cite{giannakos2024promise,denny2024desirable}. 

\citet{denny2024desirable} revealed that students prefer AI Teaching Assistants that offer debugging and code-writing support which fosters their learning autonomy, favoring guidance through problem-solving steps rather than direct solutions. However, the study also highlighted a limitation: current tools like ChatGPT do not account for a student's level of expertise, often resulting in confusing or unhelpful programming patterns.
Similarly, 
\citet{kazemitabaar2024codeaid}, noted that LLM-powered tools like ChatGPT provide immediate assistance, but by presenting a direct solution, they might discourage deep conceptual engagement. 
The authors argue that an AI assistant that is too direct might hinder skill development by bypassing critical learning opportunities, whereas overly indirect responses might overwhelm and discourage students by failing to provide adequate support. 

To achieve the right balance of support, future AI tools could utilize techniques like the Socratic method or Scaffolding-based approaches to support independent problem-solving\cite{kazemitabaar2024codeaid}.
Prior research has shown that interactive, conversational learning rooted in socio-constructivist theories—where students engage with and respond to questions from knowledgeable instructors or digital tutors—is effective in various domains~\cite {oli2023improving,tamang2021comparative}. In this context, we explore the use of LLMs to further enhance these approaches, leveraging their potential to facilitate deeper learner engagement and adaptive feedback during code-comprehension tasks. 

Code comprehension refers to the ability of an individual to understand and make sense of code written in a programming language~\cite{schulte2010introduction}. Code comprehension involves the process of reading, analyzing, and interpreting code to understand the purpose, functionality, and structure of code. Many researchers advocate for incorporating code reading into programming education, with some suggesting that students should learn to read code before they start writing it.
In today's learning environment, code comprehension has become an indispensable skill, especially as learners increasingly encounter code written by experts or generated by LLMs. However, the potential of LLMs to scaffold students' understanding of code has not been studied extensively. A critical component of effective scaffolding is the ability to identify both incomplete and incorrect elements in a student's response. In this work, we investigate how LLMs can be used to identify missing and erroneous parts in students' self-explanations of code examples, aiming to offer targeted support that enhances their understanding.

This work is part of a broader initiative to develop educational technology that improves students' understanding of code by providing tailored feedback. The system prompts students to explain their comprehension of each line of code as they read it, fostering deeper engagement. A key component of this system—and the central focus of this paper—is the automated identification of gaps and misconceptions in students' code explanations, enabling more tailored and effective feedback. Specifically, this paper explores various approaches that use LLMs to identify gaps or missing elements in learners' self-explanations of specific instructional content, such as code examples. These gaps or misconceptions are evaluated from self-explanation by correctly identifying what makes the explanation incomplete or what type of misconception is prevalent in self-explanation. 
Identifying such gaps is the major step in providing scaffolding; for instance, if a code explanation is missing a step, a tutor—whether human or AI—can offer a hint to help the learner consider and articulate the missing step. Similarly,
when a response is incorrect, the tutor needs to provide guidance to address potential misconceptions expressed by the student.
Moreover, feedback must be accurate, encouraging, and timely. It should precisely identify issues without causing confusion, and support students in developing their own solutions independently.



In this work, we investigate the use of large language models to identify missing and erroneous components in students' self-explanations of code examples.
Specifically, we investigate the following research question: 

\textbf{``Can LLMs identify gaps and misconceptions in students' code explanations? "}

We first experiment with prompting LLMs to identify the gaps and misconceptions in students' code explanations.
Although prompting LLMs in zero-shot, few-shot, chain-of-thought settings are shown to be effective across various tasks~\cite{kojima2022large},  these methods alone may not be sufficient for handling more complex or specialized tasks. Fine-tuning LLMs for more specific, well-defined tasks such as identifying gaps and misconceptions should lead to better results. We experiment with both approaches.


The following section provides a review of prior research on identifying gaps and misconceptions in students' responses, as well as the fine-tuning of LLMs for various applications. We then describe the methods and datasets used in our study and discuss the results.

\section{Related Work}

\subsection{Identifying Gaps in Student's Response}

Numerous approaches exist for automatically identifying gaps in students' responses or for identifying misconceptions in student's responses. A common method involves using engineered features to detect errors in student responses, followed by a rule-based system to provide relevant feedback or hints~\cite{kochmar2020automated,lan2015mathematical,botelho2023leveraging}. When identifying gaps, such a system compares the learner’s solution with either a rule-based or constraint-based model, highlighting where the student diverges from the correct path. Such methods are popular for their interpretability and reliability but require significant human effort to adapt to new questions and may overlook common student mistakes, leading to sub-optimal outcomes in ambiguous domains.



In the computer programming domain, researchers have introduced techniques to automatically identify gaps and misconceptions in students' responses, generating hints that offer immediate, relevant feedback to guide novices in correcting mistakes~\cite {al2023socratic}.
Recent advancements utilize large language models, employing either prompt-based techniques~\cite{al2023socratic,mcnichols2023automated,banjade2024identifying} or fine-tuning methods~\cite{jia2022insta}. 


Prior research primarily focused on assisting students with programming exercises by introducing various techniques to help novices correct mistakes and progress through the tasks. In contrast, our work emphasizes enhancing code comprehension by analyzing gaps and misconceptions in learner-generated code explanations. Similar to our approach,~\citet{banjade2024identifying} utilized zero-shot prompting of large language models to identify gaps in students' code explanations. In this work, we extend this by exploring zero-shot prompting, few-shot prompting, and fine-tuning of LLMs to identify both gaps and misconceptions in students' code explanations.

Additionally, previous studies have demonstrated that while LLMs can identify gaps and misconceptions and provide feedback to the learners, they have several limitations.
For instance,~\citet{balse2023investigating} noted that GPT-3 exhibited significant variability in the quality of generated feedback, occasionally producing incorrect and inconsistent responses. ~\citet{kiesler2023exploring} found that ChatGPT's performance varied by error type: it was effective at identifying compilation errors but struggled with logic, semantic errors, or cases involving multiple simultaneous errors in student code. 
 Similarly, ~\citet{hellas2023exploring} noted that
while GPT-3.5 could identify actual issues in student code, it had mixed success in detecting all issues and sometimes falsely identifying nonexistent issues in the code. In this study, we investigate and compare the effectiveness of prompting and fine-tuning various LLMs in accurately identifying gaps and misconceptions in students' code explanations.

\subsection{Preference Optimization using LLM}
Large Language Models have demonstrated effective instruction-following abilities, allowing them to tackle complex natural language tasks with little to no labeled data~\cite{kojima2022large}. This capability stems from their training with instruction-following data~\cite{wei2021finetuned} and reinforcement learning from human feedback 
~\citep[RLHF;][]{stiennon2020learning}.
~\citet{zhou2024lima},   
demonstrated that LLMs can effectively learn to generate specific responses through their assistant model, LIMA
achieving competitive performance with just 1,000 high-quality examples.
Instruction-tuning trains models to follow task descriptions provided in natural language, allowing them to generalize effectively to tasks they haven't encountered before~\cite{wei2021finetuned,taori2023stanford}. To further make these models more helpful and harmless, additional training with pairwise preference data is required, using techniques such as reinforcement learning with human feedback~\citep[RLHF;][]{ziegler2019fine,stiennon2020learning} or direct preference optimization~\citep[DPO;][]{rafailov2024direct}. 
Traditional methods use a reward model based on human preferences to optimize language models for tasks using RL algorithms like PPO~\cite{schulman2017proximal}, but this approach is costly and complex. 
To address this challenge, Reinforcement Learning with AI Feedback ~\citep[RLAIF;][]{lee2023rlaif} adopts rewards sourced from AI systems, such as LLMs, providing a scalable and cost-effective solution.
Optimizing LLMs' preferences with RL algorithms like PPO is effective but more complex and time-consuming than traditional supervised learning, involving tuning multiple models and real-time reward sampling.

A key improvement introduced by DPO is that the reward objective can be expressed using the optimal and reference policies, enabling model training from preference data without needing a separate reward model or policy sampling during learning~\cite{rafailov2024direct}. Hong et.al proposes Odds Ratio Preference Optimization~\cite[ORPO;][]{hong2024orpo}, a new method for training LLMs by integrating supervised fine-tuning and preference alignment into a single objective (loss function), achieving state-of-the-art results.

In the field of education,~\citet{scarlatos2024improving} proposed a feedback generation framework that optimizes the correctness and alignment with human feedback using DPO to help students solve mathematics problems. 
In the programming domain, our focus,~\citet{hicke2023chata} introduced a DPO-based approach to fine-tune LLama2 for question-answering using a dataset of Piazza\footnote{https://piazza.com/} posts from an introductory programming course. They created a proxy preference dataset from the edit history of Piazza posts, favoring final versions of answers over earlier iterations. Additionally,~\citet{kumar2024improving} proposed a method for generating Socratic questions to provide feedback on programming problems specifically to help a student debug their code. Their approach uses data augmentation techniques to synthetically generate invalid questions, which are then used to fine-tune open-source LLMs.  In contrast to these studies, which focus on debugging or code writing, our work emphasizes code comprehension by providing feedback on students' free-response explanations of code examples. We experiment with both prompting and fine-tuning LLMs using SFT and ORPO to identify gaps and misconceptions in students' self-explanations of code examples.

\section{Study Setup and Dataset:}

The Java code examples used in this work are taken from the DeepCode dataset~\cite{rus2022deepcode}, which consists of 98 annotated code examples totaling 7,157 lines of code, including comments. We selected DeepCode because it was specifically developed to enhance code comprehension skills among CS1 and CS2 students. The code examples covered the following CS concepts: \textit{ logical operators, if-else condition, loops, arrays, methods, classes and objects, exception handling, recursion, inheritance in JAVA, binary search, and sorting}. These topics were carefully selected to offer a balanced mix of complexity and variety. This dataset also includes expert-annotated code explanations designed for pedagogical purposes, such as assessment, problem-solving, and studying worked-out code examples with explanations. For our analysis, we extracted these expert annotations from the DeepCode dataset to serve as the gold standard. Additionally, we converted the Java code into Python and used GPT-4 to generate code explanations for these Python examples. Prior works have shown that LLMs can be used to generate effective code explanations~\cite{sarsa2022automatic,narayanan2024explaining}.
\\

\begin{lstlisting}[style=basicStyle1, caption={Example of java code used in the experiment}]
import java.util.ArrayList;
public class ArrayListEx1 {
    public static void main(String[] args) {
        
    ArrayList < String > travelList = new ArrayList < String >();
    
    travelList.add("Switzerland");
    travelList.add("Denmark");
    travelList.add("India");
    travelList.add("China");
    travelList.add("Thailand");
    travelList.add("Bhutan");
    
    ArrayList < String > newTravelList = new ArrayList < String >();
    for (int i = 0; i <= 5; i++) {
        
        newTravelList.add(i, travelList.get(i));
    }
        
    newTravelList.remove("China");
    newTravelList.remove(0);
    System.out.print(newTravelList);
    }
}
\end{lstlisting}
To evaluate the ability of large language models to identify gaps in code explanations, we created a dataset that varied across three key dimensions: \textit{correct, incorrect,} and \textit{incomplete} explanations. To better simulate real-world scenarios, where correct explanations may include minor errors, we augmented the dataset with correct explanations containing variations such as typos and synonyms. These variations were designed to distort the explanations slightly while preserving the underlying idea and meaning. The details of each process are explained below:

\textbf{Generating Variations of Correct Explanations:} We employed various natural language techniques such as random word deletion, synonym replacement, and random character replacement to augment the explanations by introducing typos and other modifications to simulate different quality levels of student explanations. From the original DeepCode explanations, we generated 466 different Java code explanations and 466 Python code explanations using various NLP augmentation techniques. Since the augmentations only include minor variations such as typos, we consider the explanations to be essentially correct. We annotate the gold-standard feedback for such explanations with positive feedback pointing out only minor issues in the explanations. It is important to note that the unaltered original explanations serve as our benchmark.

\begin{tcolorbox}[colback=white]
\textbf{Expert Explanation (Code shown in Listing 1):
} The program creates a person's original travel list, creates a new travel list from the original, and then modifies the new travel list. Make an arrayList travelList for holding the travel list and add destinations(countries) to the travel list. Declare an arrayList travelList whose elements are of type String and add Switzerland, Denmark, India, China, Thailand, and Bhutan. Create a list newTravelList from the original travel list travelList. Create an arrayList newTravelList for copying new travel destinations. The element of the arrayList newTravelList are of type string. Create a for loop that runs from index i= 0 to i= 5 .Get elements of arrayList travelList at position i and add the element to arrayList newTravelList at index position i.   The array elements Switzerland, Denmark, India, China, Thailand, and Bhutan are added to newTravelList. Remove China from newTravelList and remove element at index 0.  The value of newTravelList printed is [Denmark, India, Thailand, Bhutan].
\end{tcolorbox}

\textbf{Incomplete Explanation Generation} For incomplete explanation, we created simulated data inspired by the work of 
~\citet{banjade2024identifying}, where we generate the incomplete explanation by deleting two consecutive sentences from the benchmark explanation. We repeated this process to generate all the combinations of the incomplete explanation for a given benchmark explanation. This simulates a scenario where a student has gap in her code explanation, missing or overlooking important aspects of the code. These omitted details become our ground truth, which we use to evaluate whether LLMs can detect them. Our initial observation showed that consecutive pair of sentences from the generated explanation typically corresponded to an expectation, reflecting the understanding of a specific part or concept of the code. We generated a total of 1,296 incomplete code explanations, comprising 338 incomplete Java code explanations and 958 incomplete Python code explanations. 

\textbf{Incorrect Explanation Generation:}
To generate incorrect explanations, we simulated incorrect responses in a manner similar to how we simulated incomplete explanations. To simulate incorrect explanations, we first injected logical errors into the code and modified the code explanation to generate the code explanation for the error-injected version of the code. These logical errors injected were based on the misconceptions commonly observed among CS1 and CS2 students~\cite {kaczmarczyk2010identifying,ettles2018common,qian2017students}.

Some of the examples of errors injected include: constructing arrays with an off-by-one error, off-by-one error in loop conditions, replacing operator with another similar operator e.g replacing ``$>$" with ``$>=$", ignoring the 0-th index of the array, replacing floating-point division with integer division, replacing the assignment operator (=) for the comparison operator (==) and varying formatting in print statements among others. Importantly, all injected errors were logical in nature, introducing subtle variations in the code without causing it to break. The errors injected were assigned as the incorrect explanation to serve as the benchmark feedback when evaluating feedback generated using LLMs. In total, we generated 660 incorrect code explanations, consisting of 330 incorrect Java code explanations and 330 incorrect Python code explanations.

\begin{tcolorbox}[colback= white]
\textit{As shown below, we remove two consecutive sentences from expert explanation to simulate incomplete explanation} \\
\textbf{Generating Incomplete Explanation:} 

\sotwo{The program creates a person's original travel list, creates a new travel list from the original, and then modifies the new travel list. Make an arrayList travelList for holding the travel list and add destinations(countries) to the travel list.}
Declare an arrayList travelList whose elements are of type String and add Switzerland, Denmark, India, China, Thailand, and Bhutan. Create a list newTravelList from the original travel list travelList. Create an arrayList newTravelList for copying new travel destinations. The element of the arrayList newTravelList are of type string. Create a for loop that runs from index i= 0 to i= 5 .Get elements of arrayList travelList at position i and add the element to arrayList newTravelList at index position i.   The array elements Switzerland, Denmark, India, China, Thailand, and Bhutan are added to newTravelList. Remove China from newTravelList and remove element at index 0.  The value of newTravelList printed is [Denmark, India, Thailand, Bhutan].
\end{tcolorbox}

We created a \textit{dataset}\footnote{Dataset can be provided upon request} of 2,888 code explanations (1,134 in Java and 1,754 in Python) by introducing errors into incorrect explanations and simulating incomplete code examples by removing specific lines. This allowed us to analyze feedback generation using large language models. The data was split into 75\% training data, 20\% test data, and 5\% validation data. Appendix A shows examples of the simulated gaps and misconceptions in code explanations.

\section{Methodology}

We experimented under different settings: zero-shot prompting, few shot prompting, finetuning using supervised fine-tuning (SFT) and SFT with preference optimization. Each of which are explained below:

\subsection{Zero-shot Prompting}
As our baseline, we prompt LLMs to identify and generate feedback for incomplete or incorrect explanations in the simulated code explanations. We engaged in an iterative process of prompt selection, which included multiple trials and adjustments, selecting two distinct prompts for our analysis, listed below.

\begin{tcolorbox}
\textbf{\textit{P1}}: Given the following code:\{code\} and the following reference explanation: \{reference explanation\}, your task is to identify what is incorrect or missing in the following student explanation:\{student explanation\} of the code. Generate the missing part or incorrect part. If the explanation is complete and correct, aside from minor typos or issues, provide a single line of positive feedback.
\end{tcolorbox}

\begin{tcolorbox}
\textbf{\textit{P2:}} Given the following code:\{code\}, your task is to identify what is incorrect or missing in the following student explanation:\{student explanation\} of the code. Generate the missing part or incorrect part. If the explanation is complete and correct, aside from minor typos or issues, provide a single line of positive feedback.
\end{tcolorbox}

\subsection{Few-shot Prompting}

Additionally, we also investigate advanced prompting strategies, such as few-shot prompting (in-context learning), where the LLM learns from provided examples or task descriptions~\cite{brown2020language}.
We only utilized P2 in few-shot prompting to examine how LLMs can generate feedback on students' code explanations without relying on expert explanations, which are costly to produce~\cite{narayanan2024explaining}. In our few-shot setting, prompt P2 was followed by the gold standard feedback.
In few-shot learning, we utilized four types of examples: correct and complete, correct and complete with typos, incomplete, and incorrect explanations. These exemplars served as the foundation for the LLMs to learn from.

\subsection{Supervised Fine Tuning (SFT)}
In this approach, we finetuned GPT4, Llama3 and Mistral with labeled dataset. In our case, our dataset consisted of {code:\textit{code example}, explanation:{\textit{code explanation}} and feedback:\textit{feedback}}. For each code and code explanation, we used feedback as our label. 
To fine-tune OpenAI's model, we utilized the fine-tuning API provided by OpenAI.



\subsection{SFT with Preference Optimization}



In addition to SFT, preference optimization involves gathering human feedback on the model’s outputs and using this feedback as a reward signal to guide the model's behavior. By optimizing the model to produce outputs that reflect human preferences, this fine-tuning helps to reduce undesirable behaviors and ensures that responses are more socially appropriate and beneficial.

To prepare the data for the preference optimization we used the benchmark feedback as our accepted sample and the feedback generated from GPT-4 using prompts \textbf{\textit{(P1)}} and \textbf{\textit{P2}} as the rejected sample. We chose the GPT-4 generated feedback as our rejected response because, upon a cursory evaluation, we found it to be irrelevant.  The feedback from simple prompting of GPT-4 often focused on grammar, structure, and flow of the student's explanation rather than identifying gaps and misconceptions in the student's explanation, making it irrelevant in our case.


To make the feedback more helpful and relevant, we use Odds Ratio Preference Optimization (ORPO) for preference optimization. ORPO integrates an odds ratio-based penalty into the conventional negative log-likelihood loss to differentiate between the generation styles of favored and disfavored responses. Given an input sequence \( x \), the average log-likelihood of generating the output sequence \( y \), consisting of \( m \) tokens, is calculated as shown in Equation 1. The odds of generating the output sequence \( y \) given the input sequence \( x \) are defined in Equation 2.

\begin{equation}
\log P_\theta(y|x) = \frac{1}{m} \sum_{i=1}^{m} \log p_\theta(y_t \mid x, y<t)
\end{equation}

\begin{equation}
\text{odds}_\theta(y|x) = \frac{P_\theta(y|x)}{1 - P_\theta(y|x)}
\end{equation}

When \( \text{odds}_\theta(y|x) = k \), it means that the model \( \theta \) is \( k \) times more inclined to generate the sequence \( y \) than to not generate it. Equation 3 defines Odd Ratio, \( \text{OR}_\theta(y_w, y_l) \), showing how much more likely \( \theta \) is to generate \( y_w \) over \( y_l \) given input \( x \), where  \( y_w \) is the preferred response and \( y_l \) is the rejected response. The objective function for training for preference optimization has two parts: the supervised fine-tuning loss \( L_{\text{SFT}} \) and the relative log ratio loss \( L_{\text{OR}} \) as shown in in Equation 4 where \( \text{L}_{\text{SFT}} \) is the conventional causal language modeling negative log-likelihood loss and $\lambda$ represents the weighting value for \( L_{\text{OR}} \), which influences the log probablity ratio of accepted and rejected response. \( L_{\text{OR}} \) (Equation 5) maximizes the odds ratio between generating the disfavored responses \( y_w \) and \( y_l \). Further, the log sigmoid function is used to minimize \( L_{\text{OR}} \) by increasing the log odds ratio.

\begin{equation}
\text{OR}_\theta(y_w, y_l) = \frac{\text{odds}_\theta(y_w|x)}{\text{odds}_\theta(y_l|x)}
\end{equation}

\begin{equation}
\text{L}_{\text{ORPO}} = \mathbb{E}_{(x, y_w, y_l)}[\text{L}_{\text{SFT}} + \lambda \cdot \text{L}_{\text{OR}}]
\end{equation}

\begin{equation}
L_{\text{OR}}= - \log \sigma \left( \frac{\text{odds}_\theta (y_w | x)}{\text{odds}_\theta (y_l | x)} \right)
\end{equation}

\begin{table*}[h]
\centering

\begin{tabular}{ll cccc  cccc}
\toprule
&\multicolumn{4}{c}{\textbf{Python}} & \multicolumn{4}{c}{\textbf{Java}} \\
\cmidrule(lr){3-6}
\cmidrule(lr){7-10}
\textbf{Model} & \textbf{Method} & \textbf{chrF} & \textbf{METEOR} & \textbf{USE} & \textbf{BERTScr} & \textbf{chrF} & \textbf{METEOR} & \textbf{USE} & \textbf{BERTScr}  \\
\midrule

       \multirow{5}{*}{GPT4} & Prompt P1 & 0.36 & 0.22 & 0.50 & 0.84  &
       0.27 & 0.15 & 0.44 & 0.82  \\
        & Prompt P2  & 0.33 & 0.19 & 0.46 & 0.83  &
        0.26 &	0.14 &	0.42 &	0.82\\
        & Few-shot & 0.33 & 0.18 & 0.45 & 0.84 & 
        0.27 &	0.15 &	0.41 &	0.82  \\
        & SFT & \underline{0.42} & \underline{0.30} & \underline{0.53} & \underline{0.86} & 
        \underline{0.36} &	\underline{0.23} &	\underline{0.46} &	0.84  \\

\midrule
      \multirow{5}{*}{LLama3} & Prompt P1 & 0.34 & 0.20 &	0.43 &	0.84 &
      0.26 & 0.13 &	0.36 &	0.81 \\
        & Prompt P2  & 0.32 &	0.17 &	0.42 &	0.83 & 
        0.25 & 0.12 &	0.35 &	0.81  \\
        & Few-shot & 0.33 &	0.18 &	0.45 &	0.84 &	
         0.26 &	0.14 & 0.38 &	0.82 \\
        & SFT & 0.07 &	0.04 &	0.42 &	0.81 &	
        0.13 &	0.08 &	0.40 &	0.81  \\
        & ORPO & 0.28 &	0.16 &	0.44	& \underline{0.86}  &	
         0.28 &	0.16 &	0.43 &	0.85    \\
\midrule
     \multirow{5}{*}{Mistral} & Prompt P1 & 0.29 &	0.16 &	0.38 &	0.84 & 
     0.26	& 0.14 &	0.37 &	0.82  \\
        & Prompt P2  & 0.09 &	0.04 &	0.28 & 0.80 &	
        0.27 &  0.15 & 0.37 & 0.81 \\
        & Few-shot & 0.33 & 0.18 & 0.45 & 0.84 & 
        0.26 & 0.14 & 0.37 & 0.82  \\
        & SFT  & 0.32 &	0.17 &	0.44 &	0.83 &	
        0.06 &	0.03 & 0.41 &	0.81  \\
        & ORPO & 0.28 &	0.16	& 0.44 & 0.85 &	
        0.29 & 0.15 & 0.42 & \underline{0.86} 
        \\
\bottomrule
\end{tabular}
    \caption{Quantitative results of gaps and misconceptions identified in Java and Python code examples across different LLMs using various approaches.}
    \label{tab:quant_result}

\end{table*}

\subsection{Experimental Setting}
We used three distinct large language models to assess the feedback generated by LLM to identify gaps in student explanation: gpt-4-0613~\cite{openai2023gpt}, Mistral-7b-Instruct-v0.1~\cite{mixtral}, and meta-llama/Meta-Llama-3-8B ~\cite{touvron2023llama}. We accessed the open-source model for prompting and fine-tuning through Hugging Face.\footnote{https://huggingface.co} These LLMs have demonstrated state-of-the-art performance in various tasks, each utilizing different training data and algorithms, although the model size and training data specifics for the OpenAI models are not disclosed. We set the temperature parameter to 0 for all the models to ensure consistency and reproducible results. For fine-tuning we used the 8-bit quantized version of the LLama3 model and used QLora~\cite{dettmers2024qlora} to fine-tune using the parameter efficient fine-tuning  technique~\cite{peft}. For SFT, we set the maximum learning rate to 2e-4, used the AdamW optimizer, and conducted the training over 5 epochs. For fine-tuning ORPO, we set the learning rate to 8e-6, with a batch size of 2, and trained for 3 epochs using AdamW optimizer.  We fine-tuned the open-sourced model using NVIDIA A100 GPU and used FlashAttention~\cite{dao2022flashattention} to further speed up the fine-tuning process. The fine-tuned models is publicly available in hugging face\footnote{https://huggingface.co/pritiOli/OrpoLlama-3-8B-scaffolding}.

\section{Evaluation}

We evaluate the feedback generated by LLMs through prompting and fine-tuning preference optimization both qualitatively and quantitatively. 

\subsection{Quantitative Analysis}
We utilize four evaluation metrics to compare explanations generated by different sources: the character-based metric chrF~\citep{popovic2015chrf}, the word-based metric METEOR~\citep{banerjee2005meteor}, and the embedding-based metrics BERTScore~\citep{zhang2019bertscore} and Universal Sentence Encoder ~\citep[USE;][]{cer2018universal}. chrF (character n-gram F-score) measures the character-level matching between the reference text and the machine-generated text by considering both precision and recall. METEOR evaluates the similarity between words and assesses word overlap between the two texts.

BERTScore is an automated evaluation metric for text generation that assesses the similarity between candidate and reference sentences by comparing the contextual embeddings of individual tokens using cosine similarity. The Universal Sentence Encoder (USE) is a transformer-based model that converts text into high-dimensional vectors, allowing the computation of similarity between two texts based on their vector representations.
In their study, \citet{haque2022semantic} and \citet{roy2021reassessing} have noted that METEOR, chrF~\citep{popovic2015chrf}, and USE~\citep{cer2018universal} metrics better align with human preferences for code related tasks, as they assign partial credits to words. Additionally, we employ BERTScore to evaluate the generated explanations, primarily because of its extensive use as a reliable measure for assessing the faithfulness of LLMs~\citep{ji2023survey}. 

\begin{table*}[h]
\centering

\begin{tabular}{ll ccc  ccc}
\toprule
&\multicolumn{3}{c}{\textbf{Java}} & \multicolumn{3}{c}{\textbf{Python}} \\
\cmidrule(lr){3-5}
\cmidrule(lr){6-8}
\textbf{Model} & \textbf{Method} & \textbf{Correct} & \textbf{Diagnostic} & \textbf{Positive} & \textbf{Correct} & \textbf{Diagnostic} & \textbf{Positive}  \\
\midrule
       \multirow{5}{*}{GPT4} & Prompt P1 & 0.53 & 0.44 & 0.82 & 0.56 & 0.49 & 0.87   \\
        & Prompt P2   & 0.61 & 0.43 & 0.81  & 0.59 & 0.44 & 0.85 \\
        & Few-shot & 0.62 & 0.38 & 0.85  & 0.63 & 0.48 & 0.91  \\
        & SFT & 0.68  & 0.58 & \underline{0.95}  & 0.67  & 0.59 & \underline{0.97}\\
        \midrule
        
       \multirow{5}{*}{Llama3} & Prompt P1 & 0.46 & 0.28  & 0.85 & 0.44 & 0.32  & 0.87  \\
        & Prompt P2  & 0.46 & 0.28 & 0.85 & 0.50 & 0.32 & 0.88 \\
        & Few-shot & 0.42 & 0.20 & 0.88 & 0.44 & 0.30 & 0.90 \\
        & SFT &  0.56 & 0.37 & 0.68 & 0.58 & 0.48 & 0.88 \\
        & ORPO & 0.75  & 0.62 & 0.79 & \underline{0.78}  & \underline{0.67} & 0.84\\
        
        \midrule
 
       \multirow{5}{*}{Mistral} & Prompt P1 & 0.49  & 0.40 & 0.88 & 0.46  & 0.35 & 0.85  \\
        & Prompt P2  & 0.53 & 0.41 & 0.89 & 0.47 & 0.40 & 0.84 \\
        & Few-shot & 0.53 & 0.43  & 0.90 & 0.51 & 0.39  & 0.90  \\
        & SFT &  0.58 & 0.49 & 0.84 &  0.57 & 0.49 & 0.82 \\
        & ORPO & \underline{0.77}  & \underline{0.62} & 0.90 & 0.79  & 0.65 & 0.91\\
        
\bottomrule
\end{tabular}
    \caption{Qualitative results of gaps and misconceptions identified in Java and Python code examples across different LLMs using various approaches.}
    \label{tab:quanl_result}

\end{table*}

\subsection{Qualitative Analysis}
To analyze the data qualitatively, we selected a stratified sample of 220 code explanations to evaluate feedback generated by various LLMs (GPT-4, LLaMA3, Mistral) across different settings (Prompt P1 \& P2, Few-Shot Setting, SFT, ORPO). Our qualitative analysis focused on the rubrics proposed by Scarlatos et al. (2024): \textit{correct} (feedback that is accurate and relevant to both the current question and the student's response), \textit{diagnostic} (feedback that identifies errors or misconceptions in the student's answer accurately), and \textit{positive} (constructive and supportive feedback). The inter-rater agreement between 3 annotators (graduate students) is reflected by very high Cohen's Kappa coefficients, with scores of 0.94 for correctness, 0.95 for diagnostics, and 0.97 for positive metrics.


\section{Results and Discussion}

We present both quantitative and qualitative analyses of our approaches. 
Table~\ref{tab:quant_result} presents the performance outcomes of various techniques applied to large language models for identifying gaps in students' code explanations. Our findings indicate that supervised fine-tuning for GPT-4 and fine-tuning of open-source LLMs for preference alignment yield the optimal performance, achieving up to a 15\% improvement over the baseline(simple prompting).

When prompting LLMs to identify gaps in students' explanations, we observed frequent occurrences of hallucinations. This issue was particularly evident when the input data was complete and correct. In these cases, the LLMs would sometimes (27\%) generate explanations that falsely identified problems in the code explanation or even in code or focused on superficial aspects like formatting or style.
Interestingly, when dealing with code explanations that were mostly correct with minor typos or issues, the LLMs performed poorly, fabricating non-existent issues in 34\% of our qualitative analysis sample. Our findings indicate that while Large Language Models can identify incorrect and incomplete responses, they often hallucinate when prompted to identify missing gaps in explanations that are complete and correct. 

Table~\ref{tab:quanl_result} presents the findings from the qualitative analysis of prompting and fine-tuning large language models to identify gaps in students' explanations. In most cases of prompting LLMs, the feedback provided is verbose and deviant, which may not be beneficial to students if delivered directly. For example, much of the response with prompting focused on enhancing explanations, addressing typographical and clarity issues, reducing repetitive and redundant information, or providing background mathematical knowledge of the code implementation. Specifically, with LLama3, prompting the model to identify gaps in explanations mostly resulted in suggestions for fixing or optimizing code, even when it was already error-free. Additionally, LLama3 often provided positive feedback, praising the explanations even when they were incorrect or incomplete. In Table~\ref{tab:quanl_result}, we can see that fine-tuning open-source models such as LLama3 and Mistral to align to human preference significantly (up to 35\%) improves the quality of feedback generated by LLMs. Furthermore, we can observe that simple fine-tuning of GPT-4 can enhance the quality of the feedback. Although the LLMs were able to identify explanations as correct or incorrect and complete or incomplete, they struggled to pinpoint the specific missing gaps or incorrect knowledge components in the students' explanations (see Diagnostic metric in Table~\ref{tab:quanl_result}). Appendix B showcases the feedback generated by various LLMs using different methods and demonstrates how fine-tuning and preference optimization improved feedback quality.



\section{Conclusion}


In this work, we explored various methods for using LLMs to identify gaps in students' self-explanations of specific instructional material, such as explanations for code examples. Specifically, we experimented with different techniques including prompting LLMs, Supervised Fine-Tuning (SFT), and preference optimization strategies to detect gaps and misconceptions in students' self-explanations. Our findings indicate that fine-tuning approaches based on preference alignment significantly improve the quality of feedback generated. Our results show promising outcomes on employing LLMs to automatically assess and provide feedback on students' self-explanations.



\bibliography{bibliography}

\clearpage
\onecolumn

\end{document}